\documentclass[11pt,a4paper]{article}

\usepackage[T1]{fontenc} 
\usepackage{tikz}
\usepackage{tikz-cd}
\usepackage{ctable}
\usepackage{jheppub}
\usepackage{notoccite}

\begin{document}
\titlepage
\begin{flushright}
QMUL-PH-22-21\\
\end{flushright}

\vspace*{0.5cm}

\begin{center}
{\bf \Large Twisted Self-duality}

\vspace*{1cm} \textsc{
  David S. Berman\footnote{d.s.berman@qmul.ac.uk},
  Tancredi Schettini Gherardini\footnote{t.SchettiniGherardini@qmul.ac.uk} } \\

\vspace*{0.5cm} Centre for Theoretical Physics, Physical and Chemical Sciences \\
Queen Mary University of London, 327 Mile End
Road, London E1 4NS, UK\\


\vspace*{1cm}

{\bf \large Abstract}

\end{center}
We examine a generalisation of the usual self-duality equations for Yang-Mills theory when the {\it{colour space}} admits a non-trivial involution. This involution allows us to construct a non-trivial twist  which may be combined with the Hodge star to form a {\it{twisted self-dual}} curvature. We will construct a simple example of twisted self-duality for $su(2) \oplus su(2)$ gauge theory along with its explicit solutions and then dimensionally reduce from four dimensions to obtain families of non-trivial non-linear equations in lower dimensions. This  twisted self-duality constraint will be shown to arise in $E_7$ exceptional field theory  through a Scherk-Schwarz reduction and we will show how an Eguchi-Hanson gravitational instanton also obeys the twisted self-duality condition.

\vspace*{0.5cm}

\section{Introduction}
\label{sec:intro}

The importance of instanton solutions in Yang-Mills theory, described in the seminal works \cite{Belavin:1975fg, PhysRevD.14.3432, Yang:1977zf}, needs little introduction. Apart from their central role in understanding the non-perturbative physics of Yang-Mills theory, the study of self-duality has provided a fertile ground for a host of mathematical ideas such as integrabilty and topology, see for instance \cite{ATIYAH1978185,Ablowitz:2003bv}.
The central mathematical fact is that, in four dimensions with Euclidean signature, the Hodge Star acting on a two form curvature squares to one. This allows us to think of the Hodge star as an operator on the two form curvature with two eigenvalues, $+1$ and $-1$. The self dual sector has the positive eigenvalue and the anti-self dual sector has the negative eigenvalue. When we have a similar non-trivial operator acting on the colour space, i.e. an operator that squares to one, we may then also decompose the colour space in this way, according to eigenvectors of this involution. Combining the two operators of Hodge star and the internal involution allows the construction of a {\it{twisted self-duality}}, a term coined in \cite{Cremmer_1998, Cremmer_1998_2}. In fact, this sort of twisted self-duality has appeared in various formulations of supergravity and in world sheet duality symmetric formulaitons of the string theory \cite{Tseytlin:1990nb,Tseytlin:1990va,Berman:2007xn,Berman:2013eva,Alfonsi:2021bot}
To be concrete, let us denote with $\mathcal{F}^a$ the curvature two-form that takes values in some Lie algebra  
$\bf{g}$. The upper index, $a=1,...,dim(\bf{g})$ is a vector index in a particular basis for the algebra, $\bf{g}$.
Let $\mathcal{J}$ be the operator that acts on the vector space such that 
\begin{align}
    (\mathcal{J})^2=1 \, .
    \label{eq:Squares_to_one}
\end{align}
Twisted self-duality is then when the Hodge star is combined with $\mathcal{J}$, so that the curvature obeys the equation:
\begin{align}
    \mathcal{F}=*\mathcal{J}\mathcal{F} \, .
    \label{eq:TSDE_first_time}
\end{align}
We will construct an explicit (and probably the simplest) example shortly. \\
The obvious feature and a necessity for its consistency is that, the combined operator,
\begin{align}
    (*\mathcal{J})^2=1  \, .
    \label{eq:Squares_to_one}
\end{align}
This is trivially true in the case described above since both the Hodge star and $\mathcal{J}$ are involutions. But it does also beg the fascinating question of just demanding (\ref{eq:Squares_to_one}) while allowing $*^2=-1$ and $\mathcal{J}^2=-1$.
In four dimensional Lorenzian space, $*^2=-1$, which rules out real self-dual solutions. Twisted self-dual solutions however are not ruled out provided one has a twist operator such that $\mathcal{J}^2=-1$. This case is a subject of further investigation, but in this paper we focus on the Euclidean case described above.\\
It would be interesting to classify possible operators $\mathcal{J}$ for any given gauge algebra $\bf{g}$; we will not do so here. Instead, we will try to be as simple and constructive as possible. First, let us ignore Abelian groups, where the analysis is simple though certainly not uninteresting. Indeed the Abelian case which was explored as part of \cite{https://doi.org/10.48550/arxiv.1412.2768} was one of the motivations for this paper. The simplest non-abelian gauge group is $su(2)$, and so a natural way to construct an algebra with an operator $\mathcal{J}$ is simply to have two copies of $su(2)$ and have $\mathcal{J}$ map between them.
And so we consider the case of ${\bf{g}}=su(2) \oplus su(2)$ and
\begin{align}
\mathcal{J}=\begin{pmatrix}
0 & 1 \\
1 & 0
\end{pmatrix} \, .
\label{eq:J_matrix}
\end{align}
Of course, one can then use the isomorphism $so(4)=su(2) \oplus su(2)$ and think of this as an $so(4)$ gauge theory.

In what follows we will first look for solutions of this twisted self-dual equation. Then we will examine properties of the reduction of this equation to lower dimensions. This is inspired by the reduction properties of the ordinary self-dual equations and their relation to integrable systems, as described in \cite{Ablowitz:2003bv}. Following this we will show how the Eguchi-Hanson instanton originally presented in \cite{Eguchi:1979yx} can be thought of as a twisted self-dual solution.
And finally we will present the derivation of the twisted self-duality equation as a specific Scherk-Schwarz type reduction of the equations in $E_7$ exceptional field theory (for a review see \cite{Berman_2020} and references therin). This provided the original inspiration to study these twisted self-duality equations but we believe they are interesting enough in their own right that they merit study independent to the $E_7$ exceptional field theory context.

\section{Instanton Pairs}
This section is aimed at presenting the simplest example of twisted self-dual equation for a non-abelian theory, mentioned above, and constructing natural solutions by making use of the famous 't Hooft ansatz. Some exhaustive and pedagogical reviews on the subject can be found in \cite{https://doi.org/10.48550/arxiv.0802.1862, Shifman:262138, RevModPhys.51.461}.

\subsection{The Standard $SU(2)$ Instanton}
The archetype of instanton solution to Yang Mills theory is the t' Hooft ansatz constructed more than fifty years ago to solve the $SU(2)$ case in Euclidean spacetime, which can be found in \cite{PhysRevLett.37.8, PhysRevD.14.517}, for example.\\
Let us quickly review the Bogomoln’yi bound argument, which neatly shows why instantons solve the dynamics. Starting from the Yang-Mills action in Euclidean spacetime, we can complete the square to find a bound for its value:
\begin{align}
S_{Y-M} &=\frac{1}{2} \int d^{4} x \operatorname{Tr} F_{\mu \nu} F^{\mu \nu} \\
&=\frac{1}{4} \int d^{4} x \operatorname{Tr}\left(F_{\mu \nu} \mp \tilde{F}_{\mu \nu}\right)^{2} \pm 2 \operatorname{Tr} F_{\mu \nu} \tilde{F}^{\mu \nu} \geq  \frac{1}{2} \int d^{4} x \operatorname{Tr} F_{\mu \nu} \tilde{F}^{\mu \nu},
\end{align}
where $\tilde{F}_{\mu \nu}= \pm \frac{1}{2} \epsilon_{\mu \nu \rho \sigma} F^{\rho \sigma}$, with $\epsilon_{\mu \nu \rho \sigma}$ being the Levi-Civita symbol in four dimensions. The equality holds when $F= \pm \tilde{F}$.
Hence, a choice of gauge field $A_{\mu}$ such that its field strength satisfies $F_{\mu \nu} = \pm \frac{1}{2} \epsilon_{\mu \nu \rho \sigma} F_{\rho \sigma}$ minimises the action. If the sign is positive we refer to it as an instanton, while the minus sign is associated with an anti-instanton. In differential geometry terms, if we let $*$ be the Hodge star, then the statement above is compactly written as $F=\pm *F$ and $F$ said to be a (anti) self-dual curvature. The study of such objects yielded crucial insights on the nature of manifolds and their classification, see for instance the seminal work of \cite{ATIYAH1978185}. For the purpose of this article, it is only necessary to introduce the original 't Hooft ansatz for the instanton. This provides the connection and self-dual curvature on the $SU(2)$ bundle over Euclidean four-dimensional space:\\
\begin{align}
    &A_{\mu }^a= 2 \frac{\sigma_{\mu \nu}^{a}\left(x-x_{0}\right)_{\nu}}{\left(x-x_{0}\right)^{2}+\rho^{2}}, \nonumber \\
     &F_{\mu \nu }^a =2 \partial_{[\mu} A_{\nu]}{}^{a} + \epsilon^{b c a} A_{\mu}^{b} A^{c}_{\nu} =-4 \frac{\sigma_{\mu \nu}^{a}\rho^2}{[\left(x-x_{0}\right)^{2}+\rho^{2}]^2},
\end{align}
where $\sigma_{\mu \nu}^{a}$ are the self-dual 't Hooft symbols. They are defined as:
\begin{align}
     \sigma^{a \,\mu \nu} = \begin{cases}
    \sigma^{a \, 44}=0 \\
    \sigma^{a \, 4 \nu} =  - \delta^{a \nu} \hspace{0.4cm} \textrm{for} \,\,\, \nu=1,2,3 \\
    \sigma^{a \, \mu 4} =   \delta^{a \mu} \hspace{0.4cm} \textrm{for} \,\,\, \mu=1,2,3 \\
     \sigma^{a \, \mu \nu} =  \epsilon^{a \mu \nu}  \hspace{0.4cm} \textrm{for} \,\,\, \mu,\nu=1,2,3.
    \end{cases}
\end{align}
The anti self-dual curvature solution is completely analogous, just with $\sigma_{\mu \nu}^{a}$ replaced by $\bar{\sigma}_{\mu \nu}^{a}$, the anti self-dual t' Hooft symbols, which read:
\begin{align}
     \bar{\sigma}^{a \,\mu \nu} = \begin{cases}
    \bar{\sigma}^{a \, 44}=0 \\
    \bar{\sigma}^{a \, 4 \nu} =   \delta^{a \nu} \hspace{0.4cm} \textrm{for} \,\,\, \nu=1,2,3 \\
    \bar{\sigma}^{a \, \mu 4} = - \delta^{a \mu} \hspace{0.4cm} \textrm{for} \,\,\, \mu=1,2,3 \\
     \bar{\sigma}^{a \, \mu \nu} =  \epsilon^{a \mu \nu}  \hspace{0.4cm} \textrm{for} \,\,\, \mu,\nu=1,2,3.
    \end{cases}
\end{align}

\subsection{Decomposition of the twisted self-dual equation}
We will use these tools to build a solution to the the $su(2) \oplus su(2)$ twisted self-dual equation, i.e. equation \ref{eq:TSDE_first_time}, where $\mathcal{J}$ is given by \ref{eq:J_matrix}. Explicitly, this reads:
\begin{align}
    \left( \begin{array}{c}
      F  \\
     \bar{F} \\
\end{array}
\right) = * \begin{pmatrix}
0 & 1 \\
1 & 0
\end{pmatrix}
\left( \begin{array}{c}
      F  \\
     \bar{F} \\
\end{array}
\right),
\label{Twisted_self_dual_su2su2_explicit}
\end{align}
with $F,\bar{F}$ being $su(2)$ curvature forms associated with $su(2)$ connections $A,\bar{A}$, respectively.
We start by using some very basic techniques and concepts from matrix algebra: eigenvectors and eigenvalues. Let us think of $F,\bar{F}$ in terms of their colour indices, so that each of them is a three-vector in colour space. Then, we perform the following decomposition:
\begin{align}
     \left( \begin{array}{c}
      F  \\
     \Bar{F} \\
\end{array}
\right)= \left( \begin{array}{c}
      F^1 \\
      \, F^{1} \\
\end{array}
\right) + 
\left( \begin{array}{c}
      F^2  \\
      \, -F^{2} \\
\end{array}
\right),
\label{eq:Field_str_evectors_decomp}
\end{align}
where it should be clear that $F^1, F^2 $ are also three-vectors. The two vectors on the RHS are linearly independent. What we are doing is nothing but a decomposition into eigenvectors of the matrix $\begin{pmatrix}
0 & 1 \\
1 & 0
\end{pmatrix}  $.

And so, we are decomposing the $su(2) \oplus su(2)$ field strength into components that live in the two eigenspaces defined by the almost product structure above. Then, we obtain:
\begin{align}
     \left( \begin{array}{c}
      F^1 \\
      \, F^{1} \\
\end{array}
\right) + 
\left( \begin{array}{c}
      F^2  \\
     - \, F^{2} \\
\end{array}
\right) = * \begin{pmatrix}
0 & 1 \\
1 & 0
\end{pmatrix} \Bigg[  \left( \begin{array}{c}
      F^1 \\
      \, F^{1} \\
\end{array}
\right) + 
\left( \begin{array}{c}
      F^2  \\
     - \, F^{2} \\
\end{array}
\right)  \Bigg] = * \Bigg[     \left( \begin{array}{c}
      F^1 \\
      \, F^{1} \\
\end{array}
\right) -
\left( \begin{array}{c}
      F^2  \\
     - \, F^{2} \\
\end{array}
\right)   \Bigg].
\end{align}
This gives the following two equations (which are in fact six):
\begin{align}
   \, F^1= * \, F^1  , \nonumber \\
   - \, F^2 = * \, F^2.
   \label{eq:Eigenvectors_eqns}
\end{align}
Hence, we obtained two self-dual equations, which are now untwisted. We know the solution to those in terms of $F^1$ and $F^2$: they must be proportional to the 't Hooft symbols. Now, we would be tempted to complete the job by solving for the gauge fields $A^1$ and $A^2$ that correspond to $F^1$ and $F^2$, respectively. Such solutions are to the usual instanton and anti-instanton. However, the \textit{real} field strengths in our theory are $F$ and $\bar{F}$, and it is them that need to be written in terms of gauge fields $A$ and $\bar{A}$. \footnote{Clearly, if we look at the Abelian case, such consideration is superfluous. However, in the non-Abelian case, the relation between $A^{1,2}$ and $A, \bar{A}$ is highly non-trivial.}

\subsection{Solutions to the twisted self-dual equation}
A very simple solution can be found when we restrict only to one of the two eigenspaces. Specifically, setting $F^2=0$ yields the usual $su(2)$ self-dual instanton:
\begin{align}
    A_{\mu }^a=\bar{A}_{\mu }^a= A^{1}_{\mu }{}^a = 2 \frac{\sigma_{\mu \nu}^{a}\left(x-x_{0}\right)_{\nu}}{\left(x-x_{0}\right)^{2}+\rho^{2}}, \nonumber \\
     F_{\mu \nu }^a=\bar{F}_{\mu \nu}^a= F^{1}_{\mu \nu}{}^a = -4 \frac{\sigma_{\mu \nu}^{a}\rho^2}{[\left(x-x_{0}\right)^{2}+\rho^{2}]^2}.
\end{align}
This solution has clearly a "instanton number" of 2, since each of the two $su(2)$ blocks contributes with instanton number 1.\\
If we restrict to the other eigenspace, things radically change. We need to solve simultaneously:
\begin{align}
    \begin{cases}
     -F^{2}=* F^{2} \\
     F^{2}= d A - [A,A] \\
     -F^{2}= d \bar{A} - [\bar{A},\bar{A}] .
    \end{cases}
\end{align}
The first two equations are satisfied by the anti-instanton solution (i.e. instanton with instanton number $-1$) and the final equation is for a field strength with the opposite sign of the anti-instanton. Note that since the field strength is non-linear in potentials it is not possible to generate the opposite field strength by scaling the potential by $-1$. We have not be able to construct solutions for this choice.\\

Finally, we are left with the most general case of both eigenmodes contributing to the field strength. If we choose both of them to be proportional to the same scalar function $f(x^{\mu})$, then we obtain
\begin{equation}
    \begin{cases}
     F^a_{i j}= \epsilon^{aij}f(x^{\mu}) \\
     F^a_{i 4}=0
    \end{cases}
    \begin{cases}
     \bar{F}^a_{i j} = 0 \\
     \bar{F}^a_{i 4} = \delta^a_i f(x^{\mu}).
     \end{cases}
    \end{equation}
This essentially puts an electric solution in one $su(2)$ and a magnetic solution in the other $su(2)$. After making some gauge choices, it is easy to verify that a solution to such system is given by the following non-zero components of the gauge fields:
\begin{align}
 \bar{A}_{4}^a = x^a f(t) , \nonumber \\
 A^a_i = \delta^a_i \sqrt{f(t)},
\end{align}
where $f(t)$ is an arbitrary function of time. 
We will return to these solutions in section 5, in the context of $E_7$ ExFT, where the twisted self-dual equation will be accompanied by an extra constraint.

\section{Dimensional Reduction of the twisted self dual equations} 

In this section, we present some interesting dimensional reductions of the twisted self-dual equation. We first look at the case where we are dealing $2n$ copies of Maxwell theory, and find a reduction to the chirality condition of the duality symmetric sigma model studied in \cite{Tseytlin:1990va, Tseytlin:1990nb, Berman:2007xn,Kimura_2022}. Then, we look at the $su(2)\oplus su(2)$ Yang-Mills theory, providing some generalisations of the work in \cite{Ablowitz:2003bv}.

\subsection{Dimensional Reduction to the duality symmetric Sigma Model}

An $O(d,d)$ symmetric form of the string was introduced in  \cite{Tseytlin:1990va, Tseytlin:1990nb} where the idea was to double the dimensionality of the target space so that the $O(d,d)$ T-duality group would act linearly on this doubled space. Now the target space was doubled the string world sheet theory would need an additional set of  constraints to halve its degrees of freedom to the right number. This constrained doubled system should then be equivalent to the ordinary theory in the non-doubled space. This was achieved as follows. The action of the doubled sigma model with Lorentzian worldsheet reads:
\begin{align}
S=\frac{1}{4} \int_{\Sigma}\left[\mathcal{H}_{M N} \mathrm{~d} \mathbb{X}^{M} \wedge * \mathrm{~d} \mathbb{X}^{N}-\Omega_{M N} \mathrm{~d} \mathbb{X}^{M} \wedge \mathrm{d} \mathbb{X}^{N}\right] \, ,
\end{align}
where $ \mathbb{X}^{M}$ are from the world sheet perspective 2d scalars where $ (M=1..2d)$. The following twisted self-duality or Chirality relation must then be imposed:
\begin{align}
\mathrm{d} \mathbb{X}^{M}=\mathcal{J}^{M}{}_N * \mathrm{~d} \mathbb{X}^{N}.
\end{align}
The origin of the Chirality constraint from the perspective of geometric quantisation is described in \cite{Alfonsi:2021bot}.
In what follows, we will show how the twisted self-dual equation for $2n$ copies of the abelian Yang-Mills theory can be dimensionally reduced to such condition.
Let $F$ be the collection of $n$ abelian field strengths, and let $\bar{F}$ be an object of the same kind. Then, 
the twisted self-dual equation of the form \ref{Twisted_self_dual_su2su2_explicit} for the above field strengths living in Euclidean external spacetime reads:
\begin{align}
    F_{12}=\bar{F}_{34}, \quad F_{13}=\bar{F}_{42}, \quad F_{14}=\bar{F}_{23}, \nonumber \\
    \bar{F}_{12}=F_{34}, \quad \bar{F}_{13}=F_{42}, \quad \bar{F}_{14}=F_{23}.
\end{align}
We are suppressing the internal indices for brevity, and it should be understood that there are $n$ $F$'s and $n$ $\bar{F}$'s, pairwise related as above.\\
We start the dimensional reduction by setting $\partial_3 = \partial_4 =0$. Moreover, we also impose $A_3=\bar{A}_3=A_4=\bar{A}_4=0$.
This immediately sets $F_{\mu \nu}=0$ (trivially) for $(\mu,\nu) \neq (1,2)$.
Hence, we are left with $F_{12}=0=\bar{F}_{12}$ which needs to be solved for $A_1, A_2, \bar{A}_1, \bar{A}_2$. If we let $\bar{A}_2=A_1$ and $\bar{A}_1=A_2$ we obtain:
\begin{align}
    \partial_1 A_2 = \partial_2 A_1 \,\,\, , \quad
    \partial_2 A_2 = \partial_1 A_1  \, .
\end{align}
Relabelling $A_2=X$ and $A_1=\bar{X}$, and reinserting the internal structure, we get:
\begin{align}
    \partial_1 X^M = \partial_2 \bar{X}^M \,\,\, , \quad
    \partial_2 X^M = \partial_1 \bar{X}^M  ,
\end{align}
where $M=1,...,n$.
This is nothing but the equation derived in \cite{Kimura_2022} in the Born sigma model context:
\begin{align}
\mathrm{d} \mathbb{X}^{M}=\mathcal{J}^{M}{}_N * \mathrm{~d} \mathbb{X}^{N}.
\end{align}
To unpack it and show the equivalence, we note that we can write the above as:
\begin{align}
\partial_{\alpha} \mathbb{X}^{M} = \mathcal{J}^{M}{}_N \varepsilon_{\alpha \beta} \partial^{\beta} \mathbb{X}^{N},
\end{align}
where $\varepsilon_{\alpha \beta}$ is the epsilon tensor in the two-dimensional \textit{Lorentzian} worldsheet,\footnote{One might wonder that, since we are in Lorentzian space, $*^2=-1$, but we are dealing with one-forms in a two-dimensional manifold, so it does indeed square to one.} and we have:
\begin{align}
\mathcal{J}=\begin{pmatrix}
0 & 1 \\
1 & 0
\end{pmatrix}, \quad \mathbb{X}^M = \begin{pmatrix}
X^{\mu} \\
\bar{X}_{\mu}
\end{pmatrix}.
\end{align}.

Before we move to the various non-Abelian models and reductions in the next section it is worth pondering whether the four dimensional twisted self-duality might have its origin from a higher dimensional self-dual object. The natural candidate would be the M-theory five brane. The five-brane has a self-dual two form whose reduction on a torus leads to a single maxwell field and as shown in \cite{Berman:1997iz,Berman:1998va} different gauge choices in the 6 dimensional theory lead to two Maxwell fields related by a duality equation. Thinking of the 4d theory as an $u(1) \oplus u(1)$ gauge theory then the duality constraint originating from the reduciton of the fivebrane self-duality equation would be a twisted self-duality equation in 4d. This is all for the Abelian theory, having a twisted non-Abelian self-duality equation from the fivebrane is tied to the puzzle of non-Abelian Gerbes. 

\subsection{Dimensional reduction to 2D Integrable Models}
The usual (i.e. untwisted) Yang-Mills self-dual equation in Euclidean flat space is a rich source of integrable equations (see Ward conjecture, \cite{1985RSPTA.315..451W}). We first provide a quick recap on how some of its reductions have been obtained for the 4d case, following the work of \cite{Ablowitz:2003bv}. We then go on to show that the twisted case encodes all the content of the untwisted one as a special case, and emphasize how it differs with three specific reductions.
\subsection*{The Usual Self-dual Equation Case}
In its most compact form, the Yang-Mills self-dual equation reads:
\begin{align}
    F=*F,
    \label{eq:SDE}
\end{align}
where $*$ depends only on the choice of signature for a flat metric. For the Euclidean setting, which is the one that is most often examined, we have that $*^2=1$ and the equations read:
\begin{align}
    F_{12}=F_{34}, \quad F_{13}=F_{42}, \quad F_{14}=F_{23}.
\end{align}
Consider the following coordinate change:
\begin{align}
&\sigma=\frac{1}{\sqrt{2}}\left(x^{2}+i x^{3}\right), \quad \tau=\frac{1}{\sqrt{2}}\left(x^{1}-i x^{4}\right), \nonumber \\
&\tilde{\sigma}=\frac{1}{\sqrt{2}}\left(x^{2}-i x^{3}\right), \quad \tilde{\tau}=\frac{1}{\sqrt{2}}\left(x^{1}+i x^{4}\right) .
\label{eq:Change_of_coords_paper}
\end{align}
It induces a field redefinition in both $A$ and $\bar{A}$, according to:
\begin{align}
A_{4}=\frac{1}{\sqrt{2}}\left(A_{\tau}+A_{\bar{\tau}}\right), \quad A_{1}=\frac{1}{\sqrt{2}}\left(A_{\sigma}+A_{\bar{\sigma}}\right), \\
A_{2}=\frac{i}{\sqrt{2}}\left(A_{\sigma}-A_{\bar{\sigma}}\right), \quad A_{3}=-\frac{i}{\sqrt{2}}\left(A_{\tau}-A \bar{\tau}\right),
\end{align}
which leads to the simpler set of equations:
\begin{align}
    F_{\sigma \tau}=0, \quad F_{\bar{\sigma} \bar{\tau}}=0, \quad F_{\sigma \bar{\sigma}}+F_{\tau \bar{\tau}}=0.
    \label{eq:YM_simpler}
\end{align}
This system is the integrability condition for a simple isospectral linear problem, and its reductions can be obtained by simply choosing a specific dependence of the fields on the coordinates. For more details, we refer to \cite{Ablowitz:2003bv}.

\subsection*{The non-linear Schr\"odinger Equation}
With the same choice of coordinates as for the untwisted case, the twisted self-dual equation become:
\begin{align}
    & F_{\sigma \tau} + \bar{F}_{\sigma \tau}= 0 \,\,\, , \hspace{1cm} F_{\sigma \bar{\sigma}}  + \bar{F}_{ \tau \bar{\tau}} = 0 \,\,\, , \hspace{1cm}  F_{\sigma \bar{\tau}} - \bar{F}_{\sigma \bar{\tau}}  = 0 \,\,\, ,    \nonumber \\ 
    & F_{\bar{\sigma} \bar{\tau}} + \bar{F}_{\bar{\sigma} \bar{\tau}}  = 0 \,\,\, , \hspace{1cm} \bar{F}_{\sigma \bar{\sigma} } + F_{\tau \bar{\tau}}  = 0 \,\,\, , \hspace{1cm}   F_{ \bar{\sigma} \tau} - \bar{F}_{\bar{\sigma} \tau}   = 0 \,\,\, .
    \label{eq:Twistet_SD_new_coords_eucl}
\end{align}
It is easy to see that for $\bar{F}=F$, this reduces to the standard case. We now proceed to present a specific reduction, which is analogous to the one in \cite{Ablowitz:2003bv}, but for the more general case of the twisted self-dual equation.

We restrict the dependence of the fields to $x=\sigma + \bar{\sigma}$, $t=\tau$, and set $A_{\sigma}=0=A_{\bar{\sigma}}$,
and we obtain the following six equations:
\begin{align}
&\partial_{x} A_{\tau}+\partial_{x} \bar{A_{\tau}}=0 \hspace{2.65cm} \partial_{x} A_{\bar{\sigma}}-\partial_{t} \bar{A}_{\tau}- [\bar{A}_{\tau}, \bar{A}_{\bar{\tau}} ]=0 \hspace{1cm} \partial_{x} A_{\bar{\tau}}-\partial_{x} \bar{A}_{\bar{\tau}}=0 \nonumber \\
    &\partial_{x} A_{\bar{\tau}}-\partial_{t} A_{\bar{\sigma}}-\left[A_{\bar{\sigma}}, A_{\bar{\tau}} \right] + 
    \hspace{1cm}
    \partial_{x} \bar{A}_{\bar{\sigma}}-\partial_{t} A_{\tau}- \left[A_{\tau}, A_{\bar{\tau}}\right]=0 \hspace{1cm} \partial_{x} A_{\tau} - \left[A_{\bar{\sigma}}, A_{\tau}\right]-
    \nonumber \\
     &\partial_{x} \bar{A}_{\bar{\tau}}-\partial_{t} \bar{A}_{\bar{\sigma}}-\left[\bar{A}_{\bar{\sigma}}, \bar{A}_{\bar{\tau}} \right] =0 \hspace{6cm} \partial_{x} \bar{A}_{\tau}+\left[\bar{A}_{\bar{\sigma}}, \bar{A}_{\tau}\right]=0.
\end{align}
Now, we set $A_{\tau}=P$ and $\bar{A}_{\tau}=\bar{P}$, with $P,\bar{P}$ being constant matrices. We also set $A_{\bar{\tau}}=\bar{A}_{\bar{\tau}}=R$, $A_{\bar{\sigma}}=Q$ and $\bar{A}_{\bar{\sigma}}=\bar{Q}$, so that the above equations reduce to:
\begin{align}
    &\partial_{x} R-\partial_{t} Q-[Q, R] + \partial_{x} R -\partial_{t} \bar{Q} -\left[\bar{Q}, R \right] =0 , \label{eq:1} \\
    &\partial_{x} Q- [\bar{P}, R ]=0 , \label{eq:3} \\
    &\partial_{x} \bar{Q} - \left[P, R\right]=0 ,\label{eq:4} \\
    &\left[P, Q\right] = \left[\bar{P}, \bar{Q} \right] , \label{eq:5}
\end{align}
since two of them just give $0=0$.
P can be diagonalised into the form:$
\left(\begin{array}{ll}
k & 0 \\
0 &-k
\end{array}\right)$, while the other matrices explicitly read:
\begin{align}
    \bar{P}=\left(\begin{array}{ll}
l & s \\
t & -l
\end{array}\right) \,\, , \,\,\, R=\left(\begin{array}{ll}
a & b \\
c &-a
\end{array}\right) \,\, , \,\,\, Q=\left(\begin{array}{ll}
p & q \\
-r & -p
\end{array}\right),
\end{align}
with $l,s,t$ being constants. It is interesting to note that $P$ and $\bar{P}$ cannot be diagonalised simultaneously, so at this stage one of them needs to be completely undetermined.
Then, equation \ref{eq:4} immediately tells us that $\bar{Q}$ must be off-diagonal: $\bar{Q}=\left(\begin{array}{ll}
0 & \bar{q} \\
-\bar{r} &0
\end{array}\right)$, for some $\bar{q}, \bar{r}$. With this information, \ref{eq:5} gives us three constraints:
\begin{align}
    s\bar{r}= - \bar{q} t \, , \hspace{1.5cm} qk=\bar{q}l \, , \hspace{1.5cm} rk=\bar{r}l \, .
\end{align}
If we choose $s=t=0, q=\bar{q}, r=\bar{r}, k=l$, we have clearly recovered the untwisted case. Instead, we choose $s=t=0$, but leave $k$ and $l$ unrelated, with their ratio being the proportionality between $q,r$ and $\bar{q}, \bar{r}$, respectively.
This, together with equation \ref{eq:4}, gives:
\begin{align}
    b=\frac{1}{2l} \partial_x q \, , \hspace{1.5cm} c=\frac{1}{2l} \partial_x r.
\end{align}
Then, the off-diagonal entries of equation \ref{eq:3} are automatically satisfied, and the diagonal ones yield $\partial_x p =0$, which we solve by taking $p=0$. Whether expected or not, $\bar{P}$ naturally turned out to be of the same form of $P$, and the same happens here for $Q$, $\bar{Q}$. An interesting choice would be to force $P$ and $\bar{P}$ to be of a different form, we will return to this later. \\ 
Finally, we are left with \ref{eq:1}. Its diagonal part yields:
\begin{align}
    a = \frac{1}{4l} (1+ \frac{k}{l})(qr).
\end{align}
Plugging this into the diagonal part we obtain (assuming $l \neq 0$):
\begin{align}
    \partial_x^2 q - l\partial_t q +  \frac{1}{2}q(1+ \frac{k}{l})qr - k \partial_t q + \frac{k}{2l}q( 1+ \frac{k}{l})qr =0 
\end{align}
from the top-left entry, while
\begin{align}
    \partial_x^2 r + l\partial_t q + 2r \frac{1}{2}r(1+ \frac{k}{l})qr + k \partial_t r + \frac{k}{2l}r( 1+ \frac{k}{l})qr =0 
\end{align}
from the bottom-right one. \\
By setting $r=q^*$, we get:
\begin{align}
    \partial_x^2 q - (l+k) \partial_t q +  \frac{1}{2}q|q|^2(1+ \frac{k}{l}) (1 + \frac{k}{l}) =0 , \nonumber \\
     \partial_x^2 q^* + (l+k) \partial_t q^* +  \frac{1}{2}q^*|q|^2(1+ \frac{k}{l}) (1 + \frac{k}{l}) =0 ,
\end{align}
respectively.
The choice $k=l=\frac{i}{2}$ recovers the result in \cite{Ablowitz:2003bv}, while any choice of $k,l$ which satisfies $(l+k)^*=-(l+k)$ and $\big((\frac{k+l}{l})^2\big)^* = (\frac{k+l}{l})^2$ makes the two equations degenerate. The latter condition implies that only scalings by a real constant are allowed on the $|q|^2$ term, which can clearly be absorbed by a redefinition of $q$. Thus, despite the seemingly larger system of equations, the result of this reduction from the twisted self-dual equation naturally coincides with the standard (untwisted) self-dual case.

\subsection*{The Korteweg-de Vries Equation}
In the previous subsection, we assumed that $P$ was diagonalisable. When this is not the case, things radically differ. We will take $P= \left( \begin{array}{ll}
0 & 0 \\
k & 0
\end{array} \right)$, and $\bar{P}=\left( \begin{array}{ll}
0 & 0 \\
l & 0
\end{array} \right)$.\footnote{Note that in \cite{Ablowitz:2003bv}, $k$ is taken to be $1$.} Then, if $Q=\left( \begin{array}{ll}
v & z \\
w & -v
\end{array} \right)$ and $\bar{Q}=\left( \begin{array}{ll}
v' & z' \\
w' & -v'
\end{array} \right)$, equation \ref{eq:5} imposes
\begin{align}
    z'=\frac{l}{k} z  \quad \quad v' = \frac{k}{l} v.
\end{align}
Taking:
\begin{align}
   R = \left( \begin{array}{ll}
a & b \\
c & -a
\end{array} \right),
\end{align}
then equations \ref{eq:3} and \ref{eq:4} tells us that
\begin{align}
  z=z'=1 \quad \quad \partial_x v = -bl \quad \quad \partial_x w=2al \quad \quad  \partial_x v' = -bk \quad \quad \partial_x w' = 2ak ,
\end{align}
which imply also $w'=\frac{k}{l} w$. It is now the turn for equation \ref{eq:1}, which carries three independent relations. The top-right entry gives:
\begin{align}
    2 b_x - 2vb + 2a -2v'b + 2a = \nonumber \\
    -\frac{2}{l} v_{xx} + \frac{2}{l} v v_x + 2\frac{w_x}{l}  + \frac{2k}{l^2} v v_x =0,
\end{align}
from which we infer $w = v_x - \frac{1}{2}v^2(1+ \frac{k}{l})$.
Then, the diagonal part yields:
\begin{align}
    2a_x - v_t -c +bw -v'_t -c + b w' = \nonumber \\
    \frac{w_{xx}}{l} - v_t (1 + \frac{k}{l}) - \frac{v_x}{l}w(1+\frac{k}{l}) -2c =0.
\end{align}
Using the expression we found above for $w$, we get:
\begin{align}
    c= \frac{1}{2} \Big[ \frac{v_{xxx}}{l} - \frac{1}{l} (v_x^2 + v v_{xx})(1+ \frac{k}{l})  - v_t( 1+ \frac{k}{l}) - \frac{1}{l} \big(v_x^2 - \frac{1}{2} v^2 v_x (1+ \frac{k}{l} ) \big) (1+ \frac{k}{l}) \Big].
\end{align}
Then, the bottom-left entry gives the last equation:

\begin{align}
    2 c_x - w_t + 2vc -2wa - w'_t + 2v'c - 2w'a = 0.
\end{align}
Using the expressions found above, we can expand it as:
\begin{align}
    2 \Big[ \frac{v_{xxxx}}{2l} - \frac{v_x v_{xx}}{l}(1+\frac{k}{l}) - \frac{v_x v_{xx}}{2l}(1+\frac{k}{l}) - \frac{v v_{xxx}}{2l}(1+\frac{k}{l}) - \frac{v_{tx}}{2}(1+\frac{k}{l}) -  \frac{v_x v_{xx}}{l}(1+\frac{k}{l}) + \nonumber \\ + \frac{v v_x^2}{2l}(1+\frac{k}{l})^2 + \frac{v^2 v_{xx}}{4l}(1+\frac{k}{l})^2 \big] - \big(v_{xt} - v v_t (1+ \frac{k}{l}) \big) (1+ \frac{k}{l})  +
    \nonumber \\ 
    + v \Big[ \frac{v_{xxx}}{l} - \frac{1}{l} (v_x^2 + v v_{xx})(1+ \frac{k}{l})  - v_t( 1+ \frac{k}{l}) - \frac{1}{l} \big(v_x^2 - \frac{1}{2} v^2 v_x (1+ \frac{k}{l} ) \big) (1+ \frac{k}{l}) \Big] (1+ \frac{k}{l}) - \nonumber \\
    - \frac{v_x v_{xx}}{l} ( 1 + \frac{k}{l}) + \frac{v v_x^2 }{l} ( 1 + \frac{k}{l})^2 + \frac{v_{xx} v^2}{2l} ( 1 + \frac{k}{l})^2 - \frac{v_x v^3}{2l} ( 1 + \frac{k}{l})^3  
    \,\,\, .
\end{align}
Many of the terms cancel, and we are left with:
\begin{align}
    \frac{v_{xxxx}}{l} - 6 \frac{v_x v_{xx}}{l}(1 + \frac{k}{l}) -2v_{tx} (1 + \frac{k}{l}) = 0,
\end{align}
which can be rewritten as:
\begin{align}
    u_t= \frac{u_{xxx}}{2(k+l)} + 3 \frac{u u_x}{l},
\end{align}
with $u=-v_x$.
This is the KdV equation with some constant coefficients, which for $k=l=1$ trivially reduces to the standard case presented in \cite{Ablowitz:2003bv}. Even for arbitrary $k,l$, there always exist a rescaling of $x$, $t$ and $u$ which absorbs the constant coefficients. Hence, once again, the reduction of the twisted self-dual equation presented here leads to the same integrable model as its untwisted counterpart, up to rescaling.

\subsection*{The Third Option}
In the previous two subsections we studied two dimensional reductions of the twisted self-dual equation. As we pointed out in both cases, the limit in which our work reduced to the standard (untwisted) case was evident throughout our manipulations. Such limit was very straightforward because the choice $F=\bar{F}$ in \ref{Twisted_self_dual_su2su2_explicit} immediately removes the twist. And this choice was always possible in the previous two cases because we took $P$ and $\bar{P}$ to have the same form (both diagonalisable in the first one, and both lower-diagonal in the second one). Now we examine the third option: we choose $P=\left( \begin{array}{ll}
k & 0 \\
0 & -k
\end{array} \right)$, $\bar{P}=\left( \begin{array}{ll}
0 & 0 \\
\bar{k} & 0
\end{array} \right)$.\\
We start by considering equation \ref{eq:5}, with the general ansatz $Q = \left( \begin{array}{ll}
p & q \\
-r & -p
\end{array} \right)$ and $\bar{Q} = \left( \begin{array}{ll}
\bar{p} & \bar{q} \\
-\bar{r} & -\bar{p}
\end{array} \right)$. This gives 
\begin{align}
    [P,Q]= \left( \begin{array}{ll}
0 & 2qk \\
2rk & 0
\end{array} \right) = \left( \begin{array}{ll}
-\bar{q} \bar{k} & 0 \\
2\bar{k}\bar{p} & \bar{q}\bar{k}
\end{array} \right) = [\bar{P},\bar{Q}],
\end{align}
giving:
\begin{align}
    q=0, \quad \quad \bar{q}=0, \quad \quad kr=\bar{k} \bar{p} .
\end{align}
Then, taking $R=\left( \begin{array}{ll}
a & b \\
c & -a
\end{array} \right)$, \ref{eq:3} and \ref{eq:4} read:
\begin{align}
    \partial_x \left( \begin{array}{ll}
p & 0 \\
-r & -p
\end{array} \right) -  \left( \begin{array}{ll}
-b \bar{k} & 0 \\
2\bar{k} a & b \bar{k}
\end{array} \right) = 0,
\end{align}
\begin{align}
        \partial_x \left( \begin{array}{ll}
\bar{p} & 0 \\
-\bar{r} & -\bar{p}
\end{array} \right) -  \left( \begin{array}{ll}
0  & 2bk \\
-2ck  &  0
\end{array} \right) = 0.
\end{align}
This gives $b=0 \implies p=0$, \quad $\bar{p}=0 \implies r=0 \implies a=0$ \quad and \quad $\partial_x \bar{r} = 2ck$.
Hence, so far we achieved the following simplifications:
\begin{align}
    Q=\left( \begin{array}{ll}
0 & 0 \\
0 & 0
\end{array} \right), \quad \bar{Q}= \left( \begin{array}{ll}
0 & 0 \\
-\bar{r} & 0
\end{array} \right), \quad R = \left( \begin{array}{ll}
0 & 0 \\
-\frac{1}{2k} \partial_x \bar{r} & 0
\end{array} \right).
\end{align}
Plugging these in \ref{eq:1} yields:
\begin{align}
    \left( \begin{array}{ll}
0 & 0 \\
-\frac{1}{k} \partial_{xx} \bar{r} & 0
\end{array} \right) + \left( \begin{array}{ll}
0 & 0 \\ \partial_t \bar{r} & 0
\end{array} \right) = 0.
\end{align}
Hence, we obtained the diffusion equation in one-dimension (which can always be viewed as the Schr\"odinger equation in imaginary time).

\section{Eguchi-Hanson solutions as twisted self dual solutions}
In this section we move to the context of Euclidean gravity, where we found yet another appearance of the twisted self-dual equation. We present an example of a well known solution which obeys the twisted self-dual equation: the Eguchi-Hanson gravitational instanton. It was presented in \cite{Eguchi:1979yx}, and a detailed review can be found in \cite{EGUCHI197982, EGUCHI1980213}.
\subsection{Einstein-Cartan Formalism}
General Relativity in the Einstein-Cartan formalism shows evident similarities with Yang-Mills theory. Even more if the underlying spacetime is Euclidean. The basics are the following.\\
We define vierbein one-forms as:
\begin{align}
    e^a = e^a{}_{\mu} dx^{\mu} \quad \textrm{s.t.} \quad g_{\mu \nu} = \delta_{ab} e^a{}_{\mu} e^b{}_{\nu} \, .
\end{align}
Then, the connection one forms are defined by the first Cartan's structure equation:
\begin{align}
d e^{a}+\omega_{a b} \wedge e^{b}=0,
\label{eq:Cartan_1st_structure_eqn}
\end{align}
with $\omega_{a b}=-\omega_{b a}$ since they are valued in $so(4)$. Finally, the curvature reads:
\begin{align}
R_{a b} = d \omega_{a b}+\omega_{a c} \wedge \omega_{c b}.
\label{eq:Cartan_curvature}
\end{align}
It can be quickly shown that if $w$ is (anti) self-dual in the tangent space, i.e.
\begin{align}
    w_{ab}= \pm \frac{1}{2} \epsilon_{abcd} w_{cd},
\end{align}
then so is $R_{ab}$:
\begin{align}
    R_{ab}= \pm \frac{1}{2} \epsilon_{abcd} R_{cd}.
    \label{eq:Self_dual_conn_EC}
\end{align}
This equation, together with the Bianchi identity for the curvature (which holds in virtue of \ref{eq:Cartan_curvature}), implies that the Einstein equations without matter are satisfied. The solution that we are about to review is anti self-dual, i.e. it satisfies the above relation with the minus sign.

\subsection{The Eguchi and Hanson Solution}
To construct the famous instanton solution first derived by Eguchi and Hanson, it is necessary to define:
\begin{align}
&\sigma_{x}=\frac{1}{r^{2}}(x d t-t d x+y d z-z d y)=\frac{1}{2}(\sin \psi d \theta-\sin \theta \cos \psi d \phi), \nonumber \\
&\sigma_{y}=\frac{1}{r^{2}}(y d t-t d y+z d x-x d z)=\frac{1}{2}(-\cos \psi d \theta-\sin \theta \sin \psi d \phi), \nonumber \\
&\sigma_{z}=\frac{1}{r}(z d t-t d z+x d y-y d x)=\frac{1}{2}(d \psi+\cos \theta d \phi) .
\end{align}
We have that $d\sigma_x = 2 \sigma_y \wedge \sigma_z$, and cyclic permutations.
Let us take the vierbeins to be
\begin{align}
e^{a}=\left(f(r) d r, r \sigma_{X}, r \sigma_{y}, r g(r) \sigma_{z}\right),
\end{align}
with 
\begin{align}
    g(r)=\sqrt{1 - (a/r)^4} \,\,\, , \quad \quad \quad f(r) = g(r)^{-1}.
\end{align}
This gives, according to \ref{eq:Cartan_1st_structure_eqn}:
\begin{align}
&\omega^{1}{}_0=\omega^{2}{}_3=\left[1-(a / r)^{4}\right]_{x}^{\frac{1}{2}} \sigma_{x}=\left[1-(a / r)^{4}\right]^{\frac{1}{2}} e^{1} / r \, , \\
&\omega^{2}{}_0=\omega^{3}{}_1=\left[1-(a / r)^{4}\right]^{\frac{1}{2}} \sigma_{y}=\left[1-(a / r)^{4}\right]^{\frac{1}{2}} e^{2} / r  \, ,\\
&\omega^{3}{}_0=\omega^{1}{}_2=\left[1+(a / r)^{4}\right] \sigma_{z}=\left[1+(a / r)^{4}\right] e^{3} /\left(r\left[1-(a / r)^{4}\right]^{\frac{1}{2}}\right) \, .
\end{align}
The anti self-duality is really encoded in the first equality on each line. As discussed above, the resulting curvature is also anti self-dual (in the tangent space). Explicitly, it reads:
\begin{align}
&R^{1}{}_0=R^{2}{}_3 = -\frac{2 a^{4}}{r^{6}}\left(e^{1} \wedge e^{0}+e^{2} \wedge e^{3}\right) \nonumber \\
&R^{2}{}_0=R^{3}{}_1 = -\frac{2 a^{4}}{r^{6}}\left(e^{2} \wedge e^{0}+e^{3} \wedge e^{1}\right) \nonumber \\
&R^{3}{}_0=R^{1}{}_2 = +\frac{4 a^{4}}{r^{6}}\left(e^{3} \wedge e^{0}+e^{1} \wedge e^{2}\right).
\label{eq:EH_curvature_explicit}
\end{align}
Again, the anti-self-duality in the tangent space is encoded in the following relations:
\begin{align}
R^{1}{}_0=R^{2}{}_3 \, , \hspace{1.5cm} R^{2}{}_0=R^{3}{}_1\, , \hspace{1.5cm} R^{3}{}_0=R^{1}{}_2 \, .
\label{eq:Self_duality_of_EH_curvature}
\end{align}

\subsection{The twisted self-duality}
Let us now translate the twisted self-dual equation to the gravitational context. When dealing with Einstein-Cartan theory, the analogy with Yang-Mills becomes evident by identifying the curvature two-form $R$ with the field strength $F$. Hence, the twisted self-dual equation must be a relation for $R$. The next natural identification is between the internal space of Yang-Mills and the tangent space of Einstein-Cartan. The "flat" indices $a,b,etc.$ are matrix indices for elements of $so(4)$, hence any pair contains only six independent combinations. We can therefore replace any pair $ab$ with a single index $A$ running from $1$ to $6$. With these considerations, \ref{eq:Self_dual_conn_EC} can be written as:
\begin{align}
    R^A = -J^A{}_B R^B,
    \label{eq:Anti_self_dual_J}
\end{align}
with $J^A{}_B= \begin{pmatrix}
0 & 1 \\
1 & 0
\end{pmatrix}$. 
Hence, to recover the twisted self-dual equation in the context of Einstein-Cartan theory, we just need to add a Hodge star in the curved space (i.e. acting on the greek indices):
\begin{align}
    R =  J * R \iff R^A{}_{\mu \nu} = \frac{g^{1/2}}{2} \epsilon_{\mu \nu \rho \sigma} J^A{}_B R^B{}^{\rho \sigma},
    \label{eq:Self_dual_grav_inst}
\end{align}
where the determinant of the metric $g$ is different from $1$ in this case (while it was set to $1$ in the Yang-Mills context).
Now, the curvature \ref{eq:EH_curvature_explicit} already obeys \ref{eq:Anti_self_dual_J}, i.e. anti self-duality in the tangent space (see \ref{eq:Self_duality_of_EH_curvature}). This means that the action of $J$ on $R$ just introduces a minus sign. Hence, the extra requirement from the twisted self-dual equation is anti self-duality, that is
\begin{align}
    R^A{}_{\mu \nu} = - \frac{g^{1/2}}{2} \epsilon_{\mu \nu \rho \sigma} R^A{}^{\rho \sigma},
\end{align}
for $A=1,2,3$ which correspond to $(a,b)=(1,0) , (2,0) , (3,0)$, respectively.
To check it, we "flatten" the remaining two greek indices. Omitting the capital index (which goes along for the ride), we start by contracting the free indices as:
\begin{align}
    e_a{}^{\mu} e_b{}^{\nu} R_{\mu \nu} = -  e_a{}^{\mu} e_b{}^{\nu}\frac{g^{1/2}}{2} \epsilon_{\mu \nu \rho \sigma} R^{\rho \sigma},
\end{align}
where $e_a{}^{\mu}$ is given by $e_a{}^{\mu} e^a{}_{\nu} = \delta^{\mu}_{\nu}$. We use such property to bring the above equation into the form:
\begin{align}
    e_a{}^{\mu} e_b{}^{\nu} R_{\mu \nu} = - \frac{g^{1/2}}{2}  e_a{}^{\mu} e_b{}^{\nu}  e_c{}^{\tau} e_d{}^{\lambda}   \epsilon_{\mu \nu \tau \lambda} \, e^c{}_{\rho} e^d{}_{\sigma} R^{\rho \sigma},
\end{align}
Then, given $e_a{}^{\mu} e_b{}^{\nu}  e_c{}^{\tau} e_d{}^{\lambda}   \epsilon_{\mu \nu \tau \lambda} = g^{-1/2}$ and $R_{ab} \equiv e_a{}^{\mu} e_b{}^{\nu} R_{\mu \nu} $, we find:
\begin{align}
  R_{ab} = -\frac{1}{2} \epsilon_{abcd} R_{cd}.
  \label{eq:self_duality_eguchi_hansons}
\end{align}
We use the fact that $R^A= R^A{}_{ab} e^a \wedge e^b$ to read off from \ref{eq:EH_curvature_explicit}:
\begin{align}
    R^1 \propto \begin{pmatrix}
    0 & -1 & 0 & 0 \\
    1 & 0 & 0 & 0 \\
    0 & 0 & 0 & 1 \\
    0 & 0 & -1 & 0 \\
    \end{pmatrix}, \quad R^2 \propto \begin{pmatrix}
    0 & 0 & -1 & 0 \\
    0 & 0 & 0 & -1 \\
    1 & 0 & 0 & 0 \\
    0 & 1 & 0 & 0 \\
    \end{pmatrix}, 
    \quad  R^3 \propto
    \begin{pmatrix}
    0 & 0 & 0 & -1 \\
    0 & 0 & 1 & 0 \\
    0 & -1 & 0 & 0 \\
    1 & 0 & 0 & 0 \\
    \end{pmatrix}.
\end{align}
Those are the anti self-dual 't Hooft symbols (with a sign difference for $R^1$), so they indeed satisfy the anti self-duality relation.
Hence, the Eguchi-Hanson gravitation instanton is twisted self-dual in that it satisfies equation \ref{eq:Self_dual_grav_inst}.

\subsection{Topological Numbers}
Now we briefly examine the topological properties of the solution, in the light of the observations made so far. The bulk contribution to the Euler character $\chi$ is proportional to (according to Chern's formula, as described in \cite{Eguchi:1979yx}):
\begin{align}
    \int_{M} \varepsilon_{ a b c d} R^{a}{}_b \wedge R^{c}{}_d = 48 \pi^2 .
\end{align}

Now, according to the self-duality equation \ref{eq:self_duality_eguchi_hansons}, the density $\varepsilon_{ a b c d} R^{a}{}_b \wedge R^{c}{}_d$ is directly related to the density for the first Pontryagin class $p_1 \propto R^a{}_b \wedge R^b{}_a$. Specifically, we can use equation for the duality in tnagent space \ref{eq:self_duality_eguchi_hansons} to quickly compute the integral of the first Pontryagin class from the Euler class by relating the densities. This then gives:
\begin{align}
    p_1 = - \frac{1}{8\pi^2}\int_M R^a{}_b \wedge R^b{}_a =- \frac{1}{8\pi^2} \cdot 24 \pi^2  = -3 .
\end{align}
Obviously, this result agrees with the computations described in detail in \cite{EGUCHI197982}. For us it is intersting to see how the twisted self-duality equation in this example ends up relating two topological indices.

\section{Twisted self-duality in $E_7$ Exceptional Field Theory and Scherk-Schwarz reductions}
The final stage of this journey through the role of the twisted self-dual equation in physics is its appearance $E_7$ exceptional field theory. \\
We will start by providing an introduction to the theory and outlining the mechanism of Scherk-Schwarz reduction. This reduction simplifies the dynamics of the theory, making it simpler to solve. In \cite{https://doi.org/10.48550/arxiv.1412.2768}, solutions to $E_7$ exceptional field theory were found for the case where the non-abelian features are eliminated via an Abelian reduction. This section presents a continuation of the above work, since we study the simplest non-abelian generalisation of such procedure . We will show how the twisted self-dual equation emerges in the dynamics of the reduced theory, and find a particular solution.

\subsection{Introduction to $E_7$ Exceptional Field Theory}
11-dimensional supergravity is a very special theory, for many reasons. Just to mention two, 11D sugra is the maximal supergravity and it has the minimum number of dimensions required to contain the gauge group of the standard model (see \cite{Nahm:1977tg, WITTEN1981412}). However, it goes without saying that to obtain phenomenologically valid models, dimensional reduction must be performed. 
It was noticed in \cite{Cremmer:1978km} that when 11D sugra is compactified on a n -torus, $E_{n(n)}$ symmetries emerge for $n=6,7,8$.
Then, it is a natural question to ask to what extent these symmetries are present in the original theory (11D sugra before compactification). \\ The development of $E_n$ Exceptional Field Theory (ExFT) provided an answer to the above question, showing that the 11D sugra can be reformulated in a way that makes the $E_{n(n)}$ symmetry manifest prior to any compactification (for a review see \cite{Berman_2020}). \\
In this section, we are concerned with the specific case of $E_{7(7)}$ ExFT. We now briefly summarise the main aspects of such a theory, and refer the reader to the appendix for more mathematical details and to \cite{E_7_hohm_samtleben,Hull:2007zu, E_7_Pacheco_Waldram, Berman:2011jh} for a complete introduction.\\
The spacetime of $E_{7(7)}$ ExFT is $4+56$-dimensional. The first 4 coordinates $x^{\mu}$ parametrise the so-called \textit{external spacetime}, while the 56 coordinates $Y^N$ refer to the \textit{internal spacetime}. The latter ones sit in the \textrm{56} fundamental representation of $E_{7(7)}$. To recover the usual spacetime of 11D sugra, we impose the \textit{physical section condition}. This has the effect of restricting the dependence of the fields to a subset of the coordinates: the physical ones.
As a final piece of notation, we will use $\alpha$ for the index corresponding to the adjoint representation of $E_{7(7)}$, so for instance the generators of the $E_{7(7)}$ algebra are written as $(t_{\alpha})^{MN}$.
Let us now present how this construction is realised in practice.
\\
The field content of $E_{7(7)}$ ExFT is given by:
\begin{align}
    \{ g_{\mu \nu} , \mathcal{M}_{MN} , \mathcal{A}_{\mu}{}^{M}, \mathcal{B}_{\mu \nu \alpha} , \mathcal{B}_{\mu \nu M}    \}.
\end{align}
Up to a topological term, the action for this theory takes the form:
\begin{align}
S=\int \mathrm{d}^{4} x \mathrm{~d}^{56} Y e[\hat{R}+\frac{1}{48} g^{\mu \nu} \mathcal{D}_{\mu} \mathcal{M}^{M N} \mathcal{D}_{\nu} \mathcal{M}_{M N} -\frac{1}{8} \mathcal{M}_{M N} \mathcal{F}^{\mu \nu}{ }^{M} \mathcal{F}_{\mu \nu}^{N}-V\left(\mathcal{M}_{M N}, g_{\mu \nu}\right)],
\label{eq:E_7_Action}
\end{align}
where all the symbols are defined in the appendix. The key observations are that $\hat{R}$ is built out of $g_{\mu \nu}$ and $\mathcal{F}_{\mu \nu}^M$, while $\mathcal{F}_{\mu \nu}^M$ is constructed from $\mathcal{A}_{\mu}^M$.\\
The \textit{physical section condition} is given by:
\begin{align}
\left(t_{\alpha}\right)^{M N} \partial_{M} \partial_{N} \Phi=0, \quad\left(t_{\alpha}\right)^{M N} \partial_{M} \Phi \partial_{N} \Psi=0, \quad \Omega^{M N} \partial_{M} \Phi \partial_{N} \Psi=0,
\end{align}
where $ \Phi, \Psi$ are any of the fields above.
For the scope of this paper, the main observations is the following. Once the action has been varied, and the physical section condition imposed, there still a missing piece: we must impose the $E_7$ twisted self-dual equation 
\begin{align}
    \mathcal{F}_{\mu \nu}^{M}=\frac{1}{2} e \epsilon_{\mu \nu \rho \sigma} \Omega^{M N} \mathcal{M}_{N K} \mathcal{F}^{\rho \sigma K}.
    \label{eq:E_7 TSDE}
\end{align}
$\Omega^{M N}$ is the symplectic invariant form, with is invariant under $E_{7(7)}$, and:
\begin{align}
&\mathcal{F}_{\mu \nu}{ }^{M} \equiv F_{\mu \nu}{ }^{M}-12\left(t^{\alpha}\right)^{M N} \partial_{N} B_{\mu \nu \alpha}-\frac{1}{2} \Omega^{M K} B_{\mu \nu K}, \nonumber \\
&F_{\mu \nu}^{M}  \equiv 2 \partial_{[\mu} A_{\nu]}^{M}-\left[A_{\mu}, A_{\nu}\right]_{\mathrm{E}}^{M}
=2 \partial_{[\mu} A_{\nu]}^{M}-2 A_{[\mu}{ }^{K} \partial_{K} A_{\nu]}^{M}- \nonumber \\ 
&\frac{1}{2}\left(24\left(t_{\alpha}\right)^{M K}\left(t^{\alpha}\right)_{N L}-\Omega^{M K} \Omega_{N L}\right) A_{[\mu}^{N} \partial_{K} A_{\nu]}^{L}.
    \label{eq:Field_strength_ExFT_intro}
\end{align}
The reason behind the structure of $\mathcal{F}_{\mu \nu}^M$ is outlined in the appendix. \\
In recent years, progress has been made in finding solutions for this theory by eliminating completely the dependence on the external coordinates and setting the $B$-fields to zero, as described in \cite{https://doi.org/10.48550/arxiv.1412.2768}. It is easy to see that this immediately reduces $\mathcal{F}_{\mu \nu}^M$ to an Abelian field strength for 56 copies of a Maxwell gauge field $\mathcal{A}_{\mu}^M$. The aim of this section is to present a generalisation of this result to the case of $\mathcal{F}_{\mu \nu}^M$ being a non-abelian field strength of the usual Yang-Mills type. In order to obtain such a simplification of the theory, Scherk-Schwarz reduction is employed, which consists of choosing the ansatz of the schematic form $\Phi(x,Y)=U(Y)\bar{\Phi}(x)$.\footnote{This can be shown to be a consistent truncation of the theory.}\\
In the search for solutions of $E_7$ ExFT, we focus mainly on the self-duality equation, and view the equations of motion as an extra constraint.

One way of simplifying the solution to the $E_7$ ExFT dynamics is to perform a generalised Scherk-Schwarz reduction \cite{Berman:2012uy,Aldazabal:2013mya}. In this section, we will present only the details of such procedure that are relevant to our discussion.\\
We take the gauge field $\mathcal{A}_{\mu}{}^M$ to be of the form:
\begin{align}
    \mathcal{A}_{\mu}{}^M(x,Y)=\Tilde{U}^M{}_{\Bar{M}}(Y) \Bar{\mathcal{A}}_{\mu}{}^{\Bar{M}}(x),
\end{align}
where $\Tilde{U}^M{}_{\Bar{M}}(Y)= \rho^{-1}(Y) U^M{}_{\Bar{M}}(Y)$ will be referred as "twist matrix"\footnote{Note that the word \textit{twist} here has nothing to do with the operator $\mathcal{J}$ in the twisted self-dual equation.}. As usual, $Y$ denotes the internal coordinates and $x$ refers to the external ones. 
Now we make a couple of key simplifications. Firstly, we assume that the generalised Lie derivative (given in equation \ref{eq:Gen_Lie_deriv}) of one vielbein wrt to another takes the following form:
\begin{align}
   &\mathbb{L}_{\tilde{U}_{\bar{M}}} \tilde{U}^{N}{}_{\bar{N}} = \tilde{U}^K{}_{\bar{M}} \partial_{K} \tilde{U}^{N}{}_{\bar{N}}-\partial_{K} \tilde{U}^N{}_{\bar{M}} \tilde{U}^{K}{}_{\bar{N}}+\left(\lambda-\frac{1}{2}\right) \partial_{P} \tilde{U}^P{}_{\bar{M}} \tilde{U}^{N}{}_{\bar{N}} -Y^{NM}{}_{KL} \partial_{M} \tilde{U}^K{}_{\bar{M}} \tilde{U}^{L}{}_{\bar{N}} \nonumber \\ &\equiv-\tau_{\bar{M} \bar{N}}{}^{\bar{K}} \tilde{U}^{N}{}_{\bar{K}},
   \label{eq:structure_constants_generalised_lie}
\end{align}
where $\tau_{\bar{M} \bar{N}}{}^{\bar{K}}$ are constants (yet to be chosen). Secondly, we set the B-fields in the definition of the field strength (see \ref{eq:Field_strength_ExFT}) to zero.
With these choices, we can plug this ansatz in \ref{eq:Field_strength_ExFT}, which reduces to \ref{eq:Naive_field_strenght_ExFT}, and we find:
\begin{align}
    \mathcal{F}^M_{\mu \nu} = \Tilde{U}^M{}_{\Bar{M}} \big(  2\partial_{[\mu} \bar{\mathcal{A}}_{\nu]}{}^{\bar{M}} - \mathcal{A}_{[\mu}{ }^{\bar{K}} \mathcal{A}_{\nu]}{}^{\bar{P}}  \tau_{\bar{P} \bar{K}}{}^{\bar{M}} \big).
\end{align}
The full calculation is reported here just for completeness:
\begin{align}
    \mathcal{F}_{\mu \nu}^{M} 
=2 \partial_{[\mu} \mathcal{A}_{\nu]}{}^{M}-2 \mathcal{A}_{[\mu}{ }^{K} \partial_{K} \mathcal{A}_{\nu]}{}^{M}-Y^{MK}{}_{LN} \mathcal{A}_{[\mu}{}^{N} \partial_{K} \mathcal{A}_{\nu]}{}^{L} = \nonumber \\
2\Tilde{U}^M{}_{\Bar{M}} \partial_{[\mu} \bar{\mathcal{A}}_{\nu]}{}^{\bar{M}} -2 \mathcal{A}_{[\mu}{ }^{\bar{K}} \mathcal{A}_{\nu]}{}^{\bar{M}} \Tilde{U}^K{}_{\Bar{K}} \partial_K \Tilde{U}^M{}_{\Bar{M}} - Y^{MK}{}_{LN}  \mathcal{A}_{[\mu}{ }^{\bar{N}} \mathcal{A}_{\nu]}{}^{\bar{L}} \Tilde{U}^N{}_{\Bar{N}} \partial_K \Tilde{U}^L{}_{\Bar{L}} =  \nonumber  \\
2\Tilde{U}^M{}_{\Bar{M}} \partial_{[\mu} \bar{\mathcal{A}}_{\nu]}{}^{\bar{M}} - \mathcal{A}_{[\mu}{ }^{\bar{K}} \mathcal{A}_{\nu]}{}^{\bar{M}} \big(\Tilde{U}^K{}_{\Bar{K}} \partial_K \Tilde{U}^M{}_{\Bar{M}} - \Tilde{U}^K{}_{\Bar{M}} \partial_K \Tilde{U}^M{}_{\Bar{K}}\big) + \nonumber \\ 
+ \mathcal{A}_{[\mu}{ }^{\bar{K}} \mathcal{A}_{\nu]}{}^{\bar{M}} \big( - Y^{MK}{}_{LN} \Tilde{U}^N{}_{\Bar{K}} \partial_K \Tilde{U}^L{}_{\Bar{M}} \big)= \nonumber \\
2\Tilde{U}^M{}_{\Bar{M}} \partial_{[\mu} \bar{\mathcal{A}}_{\nu]}{}^{\bar{M}} + \mathcal{A}_{[\mu}{ }^{\bar{K}} \mathcal{A}_{\nu]}{}^{\bar{M}} \big( \Tilde{U}^K{}_{\Bar{M}} \partial_K \Tilde{U}^M{}_{\Bar{K}} - \Tilde{U}^K{}_{\Bar{K}} \partial_K \Tilde{U}^M{}_{\Bar{M}} - Y^{MK}{}_{LN} \Tilde{U}^N{}_{\Bar{K}} \partial_K \Tilde{U}^L{}_{\Bar{M}} \big) = \nonumber \\
2\Tilde{U}^M{}_{\Bar{M}} \partial_{[\mu} \bar{A}_{\nu]}{}^{\bar{M}} + \mathcal{A}_{[\mu}{ }^{\bar{K}} \mathcal{A}_{\nu]}{}^{\bar{M}} \big( -\tau_{\bar{M} \bar{K}}{}^{\bar{P}} \tilde{U}^{M}{}_{\bar{P}} \big) = \Tilde{U}^M{}_{\Bar{M}} \big(  2\partial_{[\mu} \bar{A}_{\nu]}{}^{\bar{M}} - \mathcal{A}_{[\mu}{ }^{\bar{K}} A_{\nu]}{}^{\bar{P}}  \tau_{\bar{P} \bar{K}}{}^{\bar{M}} \big).
\label{eq:Full_calc_field_stength}
\end{align}
Now, defining 
\begin{align}
     \mathcal{F}_{\mu \nu}^{M}= \tilde{U}^M{}_{\bar{M}}\bar{\mathcal{F}}_{\mu \nu}{ }^{\bar{M}},
\end{align}
we can write such result as:
\begin{align}
    \bar{\mathcal{F}}_{\mu \nu}{ }^{\bar{M}}=2 \partial_{[\mu} \bar{\mathcal{A}}_{\nu]}{}^{\bar{M}}+\tau_{[\bar{N} \bar{K}]}{ }^{\bar{M}} \bar{\mathcal{A}}_{\mu}{}^{\bar{N}} \bar{\mathcal{A}}_{\nu}{ }^{\bar{K}}+\ldots,
    \label{eq:Flat_field_Strength}
\end{align}
where the dots denote terms involving the extra B-fields. Thus, with the $B-$fields set to zero and the Scherk-Schwarz ansatz, we find that the "flat field strength", which is a function of the external coordinates only, recovers the usual Yang-Mills form.\\
Finally, the Scherk-Schwartz ansatz for the metric reads:
\begin{align}
    \mathcal{M}_{M N} = U_M{}^{\bar{M}} (y) U_N{}^{\bar{N}} (y)  \Bar{\mathcal{M}}_{\bar{M}\bar{N}}(x).
\end{align}   
These are all the tools needed to reduce the theory.

\subsection{Reduction of the $E_7$ Twisted Self-dual Equation}
We first look at how the above reduction is implemented on the $E_7$ twisted self-dual equation. According to \ref{eq:Squares_to_one}, we require the operator $*\Omega \mathcal{M}$ to square to one. As we anticipated in the introduction, we restrict ourselves to the case of $*^2=1$ and $\mathcal{J}^2=(\Omega \mathcal{M})^2=1$. 
\\
We present a trivial (i.e. abelian) version of the equation and a non-trivial (non-abelian) one. The abelian case is obtained by eliminating the dependence on $Y$, while the non-abelian model that we obtain involves of $su(2) \oplus su(2)$ non-abelian field strengths.\\
Also, we only focus on flat metrics. Including curved spacetimes is part of a future investigation.


We take the x-dependent part of the generalised metric to have the form
\begin{align}
    \Bar{\mathcal{M}}_{\bar{M}\bar{N}} \left(g_{m n}\right)=g^{1 / 2} \operatorname{diag}\left[g_{m n}, g^{m n, k l}, g^{-1} g^{m n}, g^{-1} g_{m n, k l}\right],
\end{align}
with the same notation as above.
Here the internal metric is constrained to be of Lorentzian signature, i.e. we take $g_{m n}= \eta_{mn} = \textrm{diag}(-1,1,1,1,1,1,1)$. By considering the possible values of the indices in $g_{m[n}g_{k]l}$, we find that for the flat case that we are focusing on, the only non-zero components are of the form:
\begin{align}
    \begin{cases}
    g_{i[i}g_{j] j} = \frac{1}{2}(1- \delta_{ij}) \hspace{0.5cm} \textrm{for}\,\,\, i,j=2,...,7 \hspace{0.5cm} \textrm{(no sum)} \\
    g_{1[1}g_{j]j}=-\frac{1}{2} \hspace{0.5cm} \textrm{(no sum)} \\
    0 \,\,\,\,\,\,\,\,\, \textrm{otherwise}.
    \end{cases}
\end{align}
Hence, we have 21 non-zero components, which we can arrange on the diagonal to form a 21x21 matrix.
Thus, we can write $\bar{\mathcal{M}}_{\bar{M} \bar{N}}$ as
\begin{align}
\left( 
\begin{array}{c | c } 
  \begin{array}{c| c c} 
     A & & \\
     \hline 
      & B & \\ 
      &  &  
  \end{array} &  \\ 
  \hline 
   &  
\begin{array}{c| c c} 
     -A & & \\
     \hline 
      & -B & \\ 
      &  & 
  \end{array}
 \end{array} 
\right) ,
\end{align}
where $A=\textrm{diag}(1,1,1,1,1,1,-1)$ and $B=\frac{1}{2}\textrm{diag}(-1,-1,-1,-1,-1,-1,1,...,1)$.
With this explicit expression for $\bar{\mathcal{M}}$, and taking $\Omega^{MN}=\begin{pmatrix}
0 & -1 \\
1 & 0
\end{pmatrix}^{MN}$ 
After performing the Scherk-Schwarz reduction, the $E_7$ twisted self-dual equation in matrix notation it reads:
\begin{align}
    \underline{\bar{\mathcal{F}}}_{\mu \nu}= \frac{1}{2} e \epsilon_{\mu \nu \rho \sigma} \begin{pmatrix}
0 & -1 \\
1 & 0
\end{pmatrix}
\left( 
\begin{array}{c | c } 
  \begin{array}{c| c c} 
     A & & \\
     \hline 
      & B & \\ 
      &  &  
  \end{array} &  \\ 
  \hline 
   &  
\begin{array}{c| c c} 
     -A & & \\
     \hline 
      & -B & \\ 
      &  & 
  \end{array}
 \end{array} 
\right) 
\underline{\bar{\mathcal{F}}}^{\rho \sigma},
\label{eq:E7_tsde_reduced}
\end{align}
Note that now the contraction of the external space (Greek) indices indices happens via the external Euclidean metric.

\subsubsection*{Vanilla case: No Dependence on Y}
The choice of the twist matrices determines the form of $\bar{\mathcal{F}}$ according to \ref{eq:structure_constants_generalised_lie} and \ref{eq:Flat_field_Strength}, so that choosing them to be constant yields the usual abelian field strength.
Then, \ref{eq:E7_tsde_reduced} gives the twisted self dual formulation of Maxwell theory, presented in \cite{Bunster_2011}. Thus, the above equation describes 28 copies of Maxwell theory.

\subsubsection*{Non-trivial case: $su(2) \oplus su(2)$}
To produce a twisted self-dual equation for $su(2) \oplus su(2)$ field strength, we need to define twist matrices that give some non-zero components of $\tau_{[\bar{N} \bar{K}]}{ }^{\bar{M}}$ (see \ref{eq:Flat_field_Strength}). We need to do this in accordance to \ref{eq:structure_constants_generalised_lie}. The choice we make is:
\begin{align}
    \tau_{\bar{N} \bar{K}}{ }^{\bar{M}}= \, \epsilon_{\bar{N} \bar{K} \bar{M}} \hspace{0.5cm} \textrm{for } \hspace{0.5cm} \bar{N} ,\bar{K}, \bar{M}, = 1,2,3 \, , 
    \label{eq:Struct_const_i}
\end{align}
and
\begin{align}
    \tau_{\bar{N} \bar{K}}{ }^{\bar{M}}= \epsilon_{\bar{N} \bar{K} \bar{M}} \hspace{0.5cm} \textrm{for } \hspace{0.5cm} \bar{N} ,\bar{K}, \bar{M}, = 29,30,31.
\end{align}
This allows us to turn off all the gauge field and  field strength components but the ones labelled by the internal index with value $1,2,3, 29,30,31$. In other words, we are left with two standard $su(2)$ field strengths, which we will denote by $F$ and $\bar{F}$, respectively. Note that we have used bars to denote quantities depending only on the external coordinates in the Scherk-Schwarz reduction, but this has nothing to do with our choice here. \\
After having performed Scherk-Schwarz reduction of the twisted-self dual equation, we end up with:
\begin{align}
    \left( \begin{array}{c}
      F  \\
     \bar{F} \\
\end{array}
\right) = * \begin{pmatrix}
0 & 1 \\
1 & 0
\end{pmatrix}
\left( \begin{array}{c}
      F  \\
     \bar{F} \\
\end{array}
\right),
\label{eq:Field_str_matrix_eqn}
\end{align}
with
\begin{align}
    F=dA - [A,A] , \hspace{0.5 cm } \bar{F}=d\bar{A} - [\bar{A},\bar{A}] , \hspace{0.5 cm} A,\bar{A} \in su(2).
\end{align}
It is understood that the Hodge star involves the Euclidean four-dimensional metric, so that this is nothing but \ref{Twisted_self_dual_su2su2_explicit}.

\subsection{Reduction of the Kinetic Term for $\mathcal{F}$}
It is evident from \ref{eq:E_7_Action} that to solve the dynamics for $\mathcal{A}_{\mu}^M$, we need to consider the equations of motion coming from the "kinetic" term $-\frac{1}{8} \mathcal{M}_{M N} \mathcal{F}^{\mu \nu}{ }^{M} \mathcal{F}_{\mu \nu}^{N}$.\footnote{The field $\mathcal{A}_{\mu}^M$ is also involved in the first term, but the equations of motion descending from it are trivially satisfied if we take the external metric to be flat.} The beauty of the Scherk-Schwarz reduction is that, as for the $E_7$ twisted self-dual equation, we can simply put bars on everything. Then, it is the specific choice of twist matrices and corresponding "structure constants" (\ref{eq:structure_constants_generalised_lie}) which determines the exact form of the equations of motion.

\subsubsection*{Vanilla case: No Dependence on Y}
For this case, the equations of motion are the same that we get for n copies of Maxwell theory. The generalised (internal) metric is what gives the trace structure to the kinetic term, so we might be worried that factors of $\pm 1$ and $\frac{1}{2}$ could make a difference. But, since the theory is abelian and $\bar{\mathcal{M}}$ is diagonal, gauge fields with different internal indices are completely uncoupled. Thus, an independent scaling of each term in the sum $-\frac{1}{8} \bar{\mathcal{M}}_{\bar{M} \bar{N}} \bar{\mathcal{F}}^{\mu \nu}{ }^{\bar{M}} \bar{\mathcal{F}}_{\mu \nu}^{\bar{N}}$ is irrelevant. The equation of motion will still be the usual
\begin{align}
    \partial_{\mu}\bar{\mathcal{F}}^{\mu \nu}=0.
    \label{eq:EoM_abelian_reduction}
\end{align}

\subsubsection*{Non-trivial case: $su(2) \oplus su(2)$}
In this case, we (consciously) decided to work with only 6 space-time gauge fields out of $\bar{\mathcal{A}}^{\bar{M}}_{\mu}$. Specifically, we chose $\bar{M}=1,2,3,29,30,31$. The blocks of $\bar{\mathcal{M}}_{\bar{M} \bar{N}}$ corresponding to $\bar{M},\bar{N}=1,2,3$ and $\bar{M},\bar{N} = 29,30,31$ are two identity matrices, which means that we have a genuine trace structure for the kinetic term in this case. Moreover, the barred and unbarred field strengths (we decompose $\mathcal{F}$ into $F, \bar{F}$, as in the previous section) are decoupled, leading to two sets of standard equations of motion for $SU(2)$ Yang-Mills:
\begin{align}
\partial_{\mu} F^{\mu \nu} -  [A_{\mu},F^{\mu \nu}] = 0 \, , \hspace{1cm} 
\partial_{\mu} \bar{F}^{\mu \nu} -  [\bar{A}_{\mu},\bar{F}^{\mu \nu}] = 0 \, .
\label{eq:Eom_exft_su(2)}
\end{align}

\subsection{The Full Dynamics}
Solutions to $E_7$ Exceptional Field Theory must satisfy both the $E_7$ twisted self-dual equation and the standard equations of motion descending from the kinetic term. Both of these have been Scherk-Schwartz reduced in the previous subsections, so that we are now in the position of proposing solutions to the (reduced) theory. Section 2 was dedicated to studying the relation \ref{eq:Field_str_matrix_eqn}, so the natural course of action is to check if the solutions developed in 2.3 also satisfy the equations of motion from the kinetic sector. It is evident that the first solution (i.e. two identical $su(2)$ instantons) does satisfy the equations of motion, since they decouple for each $su(2)$ block and each instanton individually solves the $su(2)$ Yang-Mills theory. However, a quick computation shows that, even though it has zero stress-energy tensor, the second solution does not satisfy the equations of motion. Hence, we have that the first solution solves $E_7$ exceptional field theory under the Scherk-Schwarz reduction considered, while the second one does not.

\section{Discussion}


There are a clear set remaining questions to be answered that are posed by this paper. First, as discussed at the outset, what are the possible Lorentzian equations and their solutions? This is particularly interesting since it should allow twisted self-duality where ordinary self-duality is forbidden. Is there a more interesting group that admits an appropriate involution? In terms of the exceptional field theory, can we find the full solutions once we couple to gravity including both the internal generalised metric and the external metric. Is there are interpretation in E7 ExFT of the Eguchi Hanson solution? It is tempting to see it as a gravitational instanton in the ``off-diagonal" space. Can the dimensional reductions described in the paper be useful for constructing solutions in lower dimensional ExFT's. 
Beside this set of questions inspired by ExFT it is interesting to consider other theories where twisted self-duality plays a key role. Various constructions come to mind for topological theories or generalised BF type of theories. There is much in this direction left to explore.

\section{Appendix: $E_7$ ExFT}

We now present the highlights in the construction of $E_7$ ExFT. We first sketch the structure of the "generalised" symmetry of the theory and then gauge it in order to obtain the covariant theory. Finally, we present the building blocks of the corresponding action.
For properties of $E_7$, the reader is referred to appendix B of \cite{E_7_original} and appendix B of \cite{E_7_Pacheco_Waldram}.\\
This part is mainly taken from \cite{E_7_hohm_samtleben}.\\
We define the projector onto the adjoint rep of $E_7$ as:
\begin{align}
\mathbb{P}^K{ }_{M}{}^{L}{}_N  \equiv\left(t_{\alpha}\right)_{M}^{K}\left(t^{\alpha}\right)_{N}^{L}
= \frac{1}{24} \delta_{M}^{K} \delta_{N}^{L}+\frac{1}{12} \delta_{M}^{L} \delta_{N}^{K}+\left(t_{\alpha}\right)_{M N}\left(t^{\alpha}\right)^{K L}-\frac{1}{24} \Omega_{M N} \Omega^{K L}.
\end{align}
We have that $\mathbb{P}^P{ }_{K}{}^{K}{}_P=133 $.
With this, we proceed to define the (exceptional) Lie derivative wrt the (generalised vector) $V^M$ of weight $\lambda$:
\begin{align}
    \delta V^{M}=\mathbb{L}_{\Lambda} V^{M} \equiv \Lambda^{K} \partial_{K} V^{M}-12 \mathbb{P}^M{ }_{N}{}^{K}{}_L  \partial_{K} \Lambda^{L} V^{N}+\lambda \partial_{P} \Lambda^{P} V^{M}.
\end{align}
This can be generalised straighforwardly to an arbitrary (generalised) tensor.
It follows that $\Omega^{MN}$ is invariant:
\begin{align}
    \mathbb{L}_{\Lambda} \Omega^{MN}=0.
\end{align}
Explicitly, the exceptional Lie derivative reads:
\begin{align}
&\delta_{\Lambda} V^{M}= \mathbb{L}_{\Lambda} V^{M} = \nonumber \\ 
&\Lambda^{K} \partial_{K} V^{M}-\partial_{K} \Lambda^{M} V^{K}+\left(\lambda-\frac{1}{2}\right) \partial_{P} \Lambda^{P} V^{M} -12\left(t_{\alpha}\right)^{M N}\left(t^{\alpha}\right)_{K L} \partial_{N} \Lambda^{K} V^{L}-\frac{1}{2} \Omega^{M N} \Omega_{K L} \partial_{N} \Lambda^{K} V^{L}.
\label{eq:Gen_Lie_deriv}
\end{align}
We have implicitly used the fact that $\Omega_{KL}=-\Omega_{LK}$ and $(t_{\alpha})_{KL}=(t_{\alpha})_{LK}$.
We can make this expression, and the ones that are about to follow, less cumbersome by defining the "Y-tensor":
\begin{align}
Y^{MN}{}_{KL}=12\left(t_{\alpha}\right)^{M N}\left(t^{\alpha}\right)_{K L} +\frac{1}{2} \Omega^{M N} \Omega_{K L}.
\end{align}
Then,
\begin{align}
\mathbb{L}_{\Lambda} V^{M} =
\Lambda^{K} \partial_{K} V^{M}-\partial_{K} \Lambda^{M} V^{K}+\left(\lambda-\frac{1}{2}\right) \partial_{P} \Lambda^{P} V^{M} -Y^{MN}{}_{KL} \partial_{N} \Lambda^{K} V^{L}.
\end{align}
There are some gauge parameters that give trivial transformations. They are of the form:
\begin{align}
    \Lambda^{M} \equiv\left(t^{\alpha}\right)^{M N} \partial_{N} \chi_{\alpha}, \quad \Lambda^{M}=\Omega^{M N} \chi_{N},
\end{align}
where $\chi_{N}$ satisfies the 
\begin{align}
\left(t_{\alpha}\right)^{M N} \chi_{M} \partial_{N}=\left(t_{\alpha}\right)^{M N} \chi_{M} \chi_{N}=0, \quad \Omega^{M N} \chi_{M} \partial_{N}=0.
\end{align}
We can directly verify that 
\begin{align}
    [\delta_{\Lambda_1 }, \delta_{\Lambda_2 }] = \delta_{[\delta_{\Lambda_2 }, \delta_{\Lambda_1 }]_{\mathrm{E}}},
\end{align}
where $[\cdot, \cdot]_{\mathrm{E}}$ is the E-bracket, which reads:
\begin{align}
    \left[\Lambda_{2}, \Lambda_{1}\right]_{\mathrm{E}}^{M}=2 \Lambda_{[2}^{K} \partial_{K} \Lambda_{1]}^{M}+12\left(t_{\alpha}\right)^{M N}\left(t^{\alpha}\right)_{K L} \Lambda_{[2}^{K} \partial_{N} \Lambda_{1]}^{L}-\frac{1}{4} \Omega^{M K} \Omega_{N L} \partial_{K}\left(\Lambda_{2}^{N} \Lambda_{1}^{L}\right) .
\end{align}
The Jacobiator reads (we skip many steps presented in \cite{E_7_hohm_samtleben})
\begin{align}
J^{M}\left(V_{1}, V_{2}, V_{3}\right)=&-\frac{1}{2}\left(t_{\alpha}\right)^{M K} \partial_{K}\left(\left(t^{\alpha}\right)_{P L}\left(V_{1}^{P}\left[V_{2}, V_{3}\right]_{\mathrm{E}}^{L}+\text { cycl. }\right)\right) \\
&+\frac{1}{12} \Omega^{M K} \Omega_{N L}\left(V_{1}^{N} \partial_{K}\left[V_{2}, V_{3}\right]_{\mathrm{E}}^{L}+\left[V_{1}, V_{2}\right]_{\mathrm{E}}^{N} \partial_{K} V_{3}^{L}+\mathrm{cycl} .\right).
\end{align}
So, it is trivial in the sense that it does not generate any transformation.
We can now construct a theory that is manifestly invariant under the generalised diffeomorphisms defined above. To do that, we introduce gauge fields $\mathcal{A}_{\mu}^M$ and define the associated covariant derivative as:
\begin{align}
    \mathcal{D}_{\mu} \equiv \partial_{\mu} - \mathbb{L}_{\mathcal{A}_{\mu}}.
\end{align}
Then, it follows that:
\begin{align}
\mathcal{D}_{\mu} V^{M} \equiv  D_{\mu} V^{M}-\lambda \partial_{K} \mathcal{A}_{\mu}{ }^{K} V^{M} 
\equiv  \partial_{\mu} V^{M}- \mathcal{A}_{\mu}{ }^{K} \partial_{K} V^{M}+V^{K} \partial_{K} \mathcal{A}_{\mu}{ }^{M}+\frac{1-2 \lambda}{2} \partial_{K} \mathcal{A}_{\mu}{ }^{K} &V^{M} \nonumber \\
+12\left(t_{\alpha}\right)^{M N}\left(t^{\alpha}\right)_{K L} \partial_{N} \mathcal{A}_{\mu}{ }^{K} V^{L}+\frac{1}{2} \Omega^{M N} \Omega_{K L} \partial_{N} \mathcal{A}_{\mu}{ }^{K} &V^{L},
\end{align}
where we implicitly defined $D_{\mu} V^{M}$. 
If we require the correct transformation for the covariant derivative, we find:
\begin{align}
&\delta \mathcal{A}_{\mu}^{M}= \partial_{\mu} \Lambda^{M}-\mathcal{A}_{\mu}^{K} \partial_{K} \Lambda^{M}+\Lambda^{K} \partial_{K} \mathcal{A}_{\mu}^{M} +12\left(t_{\alpha}\right)^{M N}\left(t^{\alpha}\right)_{K L} \Lambda^{L} \partial_{N} \mathcal{A}_{\mu}{ }^{K}+\frac{1}{2} \Omega^{M N} \Omega_{K L} \Lambda^{L} \partial_{N} \mathcal{A}_{\mu}^{K} \nonumber\\
&= D_{\mu} \Lambda^{M}-\frac{1}{2}\left(\partial_{K} A_{\mu}^{K}\right) \Lambda^{M} \equiv \mathcal{D}_{\mu} \Lambda^{M}.
\end{align}
Hence, we find that $ \Lambda^{M}$ has weight $\lambda= \frac{1}{2}$.\\
It is now the turn for the field strength:
\begin{align}
F_{\mu \nu}^{M}  \equiv 2 \partial_{[\mu} \mathcal{A}_{\nu]}^{M}-\left[\mathcal{A}_{\mu}, \mathcal{A}_{\nu}\right]_{\mathrm{E}}^{M}
=2 \partial_{[\mu} \mathcal{A}_{\nu]}^{M}-2 \mathcal{A}_{[\mu}{ }^{K} \partial_{K} \mathcal{A}_{\nu]}^{M}-\overbrace{\frac{1}{2}\left(24\left(t_{\alpha}\right)^{M K}\left(t^{\alpha}\right)_{N L}-\Omega^{M K} \Omega_{N L}\right)}^{Y^{MK}{}_{LN}} \mathcal{A}_{[\mu}^{N} \partial_{K} \mathcal{A}_{\nu]}^{L}.
\label{eq:Naive_field_strenght_ExFT}
\end{align}
This is not covariant wrt generalised diffeomorphisms (and this is a consequence of the non-vanishing Jacobiator, see \cite{E_7_hohm_samtleben}). The variation is given by:
\begin{align}
\delta F_{\mu \nu}{ }^{M}= 2 D_{[\mu} \delta \mathcal{A}_{\nu]}^{M}-\partial_{K} \mathcal{A}_{[\mu}{ }^{K} \delta \mathcal{A}_{\nu]}^{M}-12\left(t_{\alpha}\right)^{M K}\left(t^{\alpha}\right)_{N L} \partial_{K}\left(A_{[\mu}^{N} \delta \mathcal{A}_{\nu]}^{L}\right) -\nonumber \\ \frac{1}{2} \Omega^{M K} \Omega_{L N}\left(\mathcal{A}_{[\mu}^{N} \partial_{K} \delta \mathcal{A}_{\nu]}^{L}-\partial_{K} \mathcal{A}_{[\mu}^{N} \delta \mathcal{A}_{\nu]}^{L}\right).
\end{align}
We recover covariance by including two additional fields, $B_{\mu \nu \alpha}$ and $B_{\mu \nu K}$, in the definition of the field strength:
\begin{align}
    \mathcal{F}_{\mu \nu}{ }^{M} \equiv F_{\mu \nu}{ }^{M}-12\left(t^{\alpha}\right)^{M N} \partial_{N} B_{\mu \nu \alpha}-\frac{1}{2} \Omega^{M K} B_{\mu \nu K}.
    \label{eq:Field_strength_ExFT}
\end{align}
This works as follows. The variation of the extended field strength reads:
\begin{align}
    \delta \mathcal{F}_{\mu \nu}^{M}=2 \mathcal{D}_{[\mu} \delta A_{\nu]}^{M}-12\left(t^{\alpha}\right)^{M N} \partial_{N} \Delta B_{\mu \nu \alpha}-\frac{1}{2} \Omega^{M K} \Delta B_{\mu \nu K},
\end{align}
with
\begin{align}
\Delta B_{\mu \nu \alpha} &\equiv \delta B_{\mu \nu \alpha}+\left(t_{\alpha}\right)_{K L} \mathcal{A}_{[\mu}^{K} \delta \mathcal{A}_{\nu]}^{L} \\
\Delta B_{\mu \nu K} &\equiv \delta B_{\mu \nu K}+\Omega_{L N}\left(A_{[\mu}^{N} \partial_{K} \delta \mathcal{A}_{\nu]}^{L}-\partial_{K} \mathcal{A}_{[\mu}^{N} \delta \mathcal{A}_{\nu}]^{L}\right).
\end{align}
Then, with the choices
\begin{align}
\delta_{\Lambda} \mathcal{A}_{\mu}^{M} &=\mathcal{D}_{\mu} \Lambda^{M} \\
\Delta_{\Lambda} B_{\mu \nu \alpha} &=\left(t_{\alpha}\right)_{K L} \Lambda^{K} \mathcal{F}_{\mu \nu}^{L} \\
\Delta_{\Lambda} B_{\mu \nu M} &=-\Omega_{K L}\left(\mathcal{F}_{\mu \nu}{ }^{K} \partial_{M} \Lambda^{L}-\Lambda^{L} \partial_{M} \mathcal{F}_{\mu \nu}{ }^{K}\right)
\end{align}
we obtain the covariant transformation of the field strength:
\begin{align}
    \delta_{\Lambda} \mathcal{F}_{\mu \nu}^{M}=\Lambda^{K} \partial_{K} \mathcal{F}_{\mu \nu}^{M}-12 \mathbb{P}^{M}{ }_{N}{ }^{K}{ }_{L} \partial_{K} \Lambda^{L} \mathcal{F}_{\mu \nu}{ }^{N}+\frac{1}{2} \partial_{K} \Lambda^{K} \mathcal{F}_{\mu \nu}{ }^{M}.
\end{align}
Finally, we now give explicit expressions for the terms that appear in the action \ref{eq:E_7_Action}. We have:
\begin{align}
\widehat{R}_{\mu \nu}^{a b} &\equiv R_{\mu \nu}^{a b}[\omega]+\mathcal{F}_{\mu \nu}^{M} e^{a \rho} \partial_{M} e_{\rho}^{b}, \nonumber \\
V=&-\frac{1}{48} \mathcal{M}^{M N} \partial_{M} \mathcal{M}^{K L} \partial_{N} \mathcal{M}_{K L}+\frac{1}{2} \mathcal{M}^{M N} \partial_{M} \mathcal{M}^{K L} \partial_{L} \mathcal{M}_{N K}, \nonumber \\
&-\frac{1}{2} g^{-1} \partial_{M} g \partial_{N} \mathcal{M}^{M N}-\frac{1}{4} \mathcal{M}^{M N} g^{-1} \partial_{M} g g^{-1} \partial_{N} g-\frac{1}{4} \mathcal{M}^{M N} \partial_{M} g^{\mu \nu} \partial_{N} g_{\mu \nu}.
\end{align}
where $ R_{\mu \nu}^{a b}$ is the spin connection for the vierbein $e^{a \mu}$ and $g=e^2 =\textrm{det}(e_a{}^{\mu})^2=\textrm{det}(g_{\mu \nu})$. We do not want to dwell too much on details that are not relevant for our discussion, so we refer the reader to \cite{E_7_hohm_samtleben} for more details on the topological term and for why the action takes this specific form.
\section*{Acknowledgements}
The authors would like to thank Costis Papageorgakis, Malcom Perry and Chris White for discussions and comments on the manuscript and in particular Rod Halburd for early discussions on self-duality and dimensional reduction. 
TSG would like to thank Emma Albertini, Pietro Capuozzo and Emanuele Panella for their invaluable suggestions. TSG was supported by an STFC studentship. DSB thanks Pierre Andurand for his generous donation supporting this research.

\bibliographystyle{unsrtnat}

\bibliography{biblio}{}

\end{document}